# Low temperature magnetism of gold nano particles contained in electrochemical sugar recognition system

T. Goto[1], Y. Kitamoto[1], K. Matsui[1,2], H. Kuroe[1], A. Endo[1], T. Hashimoto[1], T. Hayashita[1], S. Iguchi[3], T. Sasaki[3]

[1]Faculty of Science and Engineering, Sophia University, Chiyodaku Tokyo, 102-8554 Japan
[2]The Institute for Solid State Physics, The University of Tokyo, Kashiwanoha, Kashiwa, Chiba 277-8581 Japan
[3]Institute for Material Research, Tohoku University, Katahira, Aoba-ku, Sendai 980-8577, Japan

**Low temperature magnetic properties were investigated on the gold nano particles (GNP) with an average size of 11.5 nm, assembled with molecules of ruthenium complex (Ru0) and phenylboronic acid (B0) by the proton nuclear resonance ($^{1}$H-NMR) and susceptibility measurements. The temperature dependence of the NMR shift and the uniform susceptibility was described as the sum of the Curie-Weiss term and a positive constant term. From the former, the average number of Ru0 on each GNP was estimated to be 118, which is 23 % of the calculation. The finite positive constant term shows a clear contrast with the well-known fact that the bulk gold is diamagnetic. Finally, a disappearance of the motional narrowing effect in the proton NMR spectra below 60 K indicates that the wavering motion of Ru0 complexes on GNP at room temperature is frozen at low temperatures.**

## I. Introduction

The recently reported electrochemical sugar recognition system consisting of a gold nano particle (GNP), a ruthenium complex and a pheylboronic acids, attracts much interest because of its high sensitivity for various sugars such as D-glucose or D-fructose [1]. The schematic of tris (acetylacetonato) ruthenium (Ru0) and phenylboronic acid (B0) assembled on a GNP is shown in Fig. 1. In addition to its interesting and practical function, from a viewpoint physics, this system can be considered as an $S$ = 1/2 spin cluster (Ru0's) assembled on a nano-sized metal particle (GNP). The purpose of this study is to investigate its magnetic properties; the result will be of help for understanding this nano-sized gold magnet [2-4] as well as for elucidating the mechanism of its sugar-sensing function.

## II. Experimental

The gold nano particles (GNPs) were prepared from $HAuCl_4$ by the reduction of sodium citrate following the general method [1,5,6]. The average diameter of the GNPs was determined approximately by the dynamic light scattering (DLS) to be 11.5 nm (PLD ≃0.2) [7]. They are assembled with ruthenium complex (Ru0) [8] or phenylboronic acid (B0) [9]; for the present study, the three samples of powder form were prepared, Ru0/B0/GNP, Ru0/GNP and standalone GNP. The attachment of Ru0 and B0 was confirmed by the form of reduction wave in the electrochemical measurements [1]. We have obtained dried powder sample by the following method. First, the copolymer of copolymers [poly($N$-n-isopropylacrylamide-*co*-acryloyldiethyletriamine)] is attached to the surface of GNPs, which, then, are assembled with Ru0 and B0 in ethanol. These composites are heated and then cooled, and at this process, each particle is completely separated. Finally, they are centrifuged and dried for the succeeding measurements.

The uniform susceptibility of dried powder [10-12] was measured by conventional SQUID magnetometer (Quantum Design Co. Ltd.) under the magnetic field of 1T in the temperature range 2 and 280 K. In measuring the susceptibility, the powder samples were wrapped with thin mylar sheet (Saranwrap, Asahikasei Co., Ltd.). The susceptibility of the sheet alone was separately measured (see the inset of Fig. 2), and subtracted as a background from the sample data.

The $^{1}$H-NMR spectra were obtained in the temperature range between 1.8 and 300 K either by Fourier-transform of Free-induction decay (FID) signal, or plotting the spin-echo amplitude against the applied field [13].

## III. Results and Discussion

Figure 2 shows the temperature dependence of the uniform susceptibility of Ru0/B0/GNP, from which the background of the wrapping film (inset) was subtracted. It was described by sum of the Curie-Weiss term and constant term as $C/(T+\Theta)+\chi_0$. As for the other two samples of Ru0/GNP and GNP, the uniform susceptibility was measured and analyzed by the same procedures as shown in Fig. 3. Each parameter was obtained by fitting, which was performed at the temperature region above 10 K, so that the effect of magnetization saturation was avoided. Obtained parameters are shown in Table 1. For the sample of standalone GNP, $C$ was zero within the experimental precision.

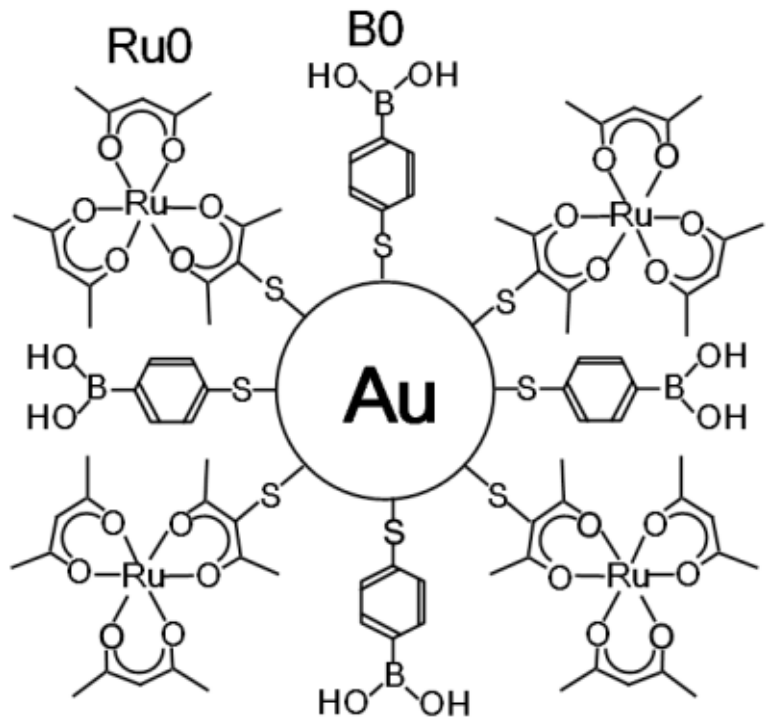

**FIG. 1** Schematic of molecular structure of a gold nanoparticle (GNP), Ru-based complexes (Ru0) and boron acid molecules (B0).

The observed Curie-Weiss term is immediately assigned to the localized $S = 1/2$ spin on each trivalent ruthenium ion. The monotonic increase in the susceptibility without showing anomalies down to the lowest temperatures indicates the absence of a long-range magnetic ordering. The smallness of $\Theta$ indicates that those spins are only weakly coupled with one another and can be considered nearly free spins. Utilizing this fact with the obtained parameters in Table 1, we first try to estimate $n_{\mathrm{Ru}}$, the average number of Ru0 complex assembled on each GNP. Generally, the uniform susceptibility of $S = 1/2$ free spins is given as $n_{\mathrm{Ru}} g^2 \mu_{\mathrm{B}}^2 S(S+1)/3k_{\mathrm{B}}T$, where $g$, $\mu_{\mathrm{B}}$ and $k_{\mathrm{B}}$ denote $g$-factor, the Bohr magneton and the Boltzmann constant, respectively. If we assume $g \simeq 2$, and use the mass of a GNP with the mean diameter of 11.5 nm, $m_{\mathrm{GNP}} \simeq 1.54 \times 10^{-17}$ (g), one can simply estimate $n_{\mathrm{Ru}} \simeq C m_{\mathrm{GNP}} k_{\mathrm{B}}/\mu_{\mathrm{B}}^2$ to be 276 and 118 for Ru0/GNP and Ru0/B0/GNP, respectively. This estimation of $n_{\mathrm{Ru}}$ is much smaller than the value of 513, calculated with a simple calculation assuming the closed-packed structure. That is, we assumed that the shape of head part of Ru0 is sphere-like, and that those spheres are assembled on the surface of GNP in the closed-packed way. For the sensitivity and selectivity of the sugar recognition system depends delicately on $n_{\mathrm{Ru}}$, the present estimation of $n_{\mathrm{Ru}}$ will be a direct help to the improvement. In order to pursue further precision, the determination of GNP size by TEM is necessary, which is now on the progress.

Next, in Fig. 4, the Fourier spectrum of $^1$H-NMR FID signal for Ru0/B0/GNP at room temperature is shown. It consists of two extremely sharp peaks separated by 4.5 Oe and a broad tails of 15 Oe in the definition of full width at half maximum (FWHM). The appearance of this fine structure indicates the motional narrowing effect due to the weaving motion of Ru0's. With decreasing temperature, this FID signal disappeared below 60 K, and instead, appeared a spin-echo signal. Figure 5 shows the profile of spectra at various temperatures. Note that the fine structure observed at the room temperature is completely lost below 60 K, demonstrating that the weaving motion of Ru0 is frozen and hence the motional narrowing effect.

| Sample | $\chi_0$ ($10^{-7}$×emu/g) | C ($10^{-7}$×K·emu/g) | Θ (K) |
|---|---|---|---|
| GNP | 7.7(1) | none | none |
| Ru0/GNP | 1.50(5) | 111(10) | 0.6(3) |
| Ru0/B0/GNP | 0.50(5) | 47.3(10) | 0.4(2) |

**TABLE 1** Magnetic parameters for the three samples obtained by magnetization measurements.

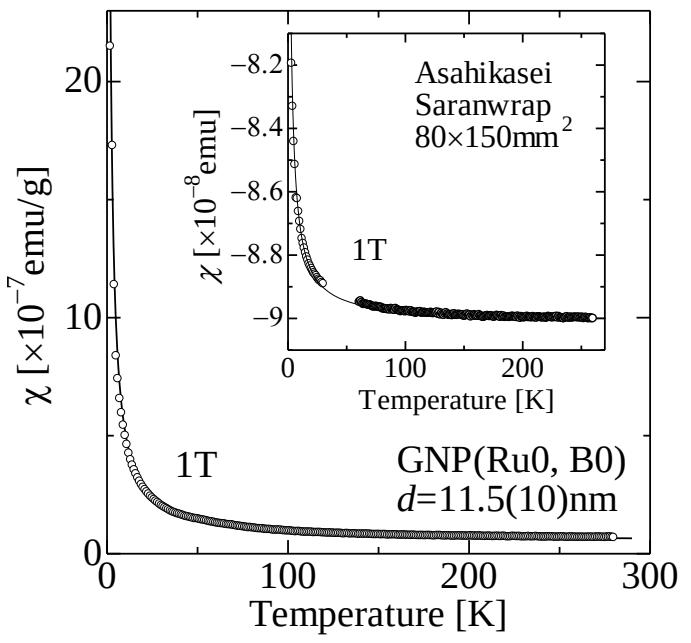

**FIG. 2** Temperature dependence of the uniform susceptibility of GNP with Ru0 and B0. The inset shows the back ground.

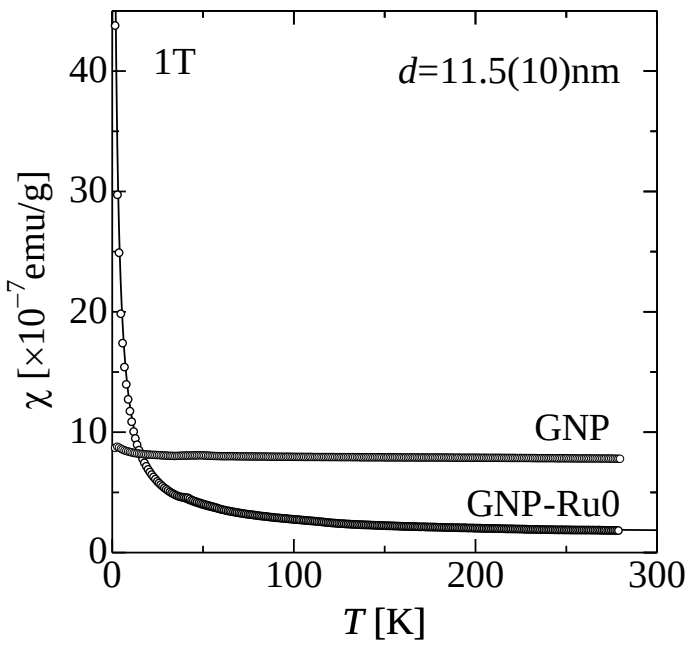

**FIG. 3** Temperature dependence of the uniform susceptibility of Ru0/GNP and GNP alone.

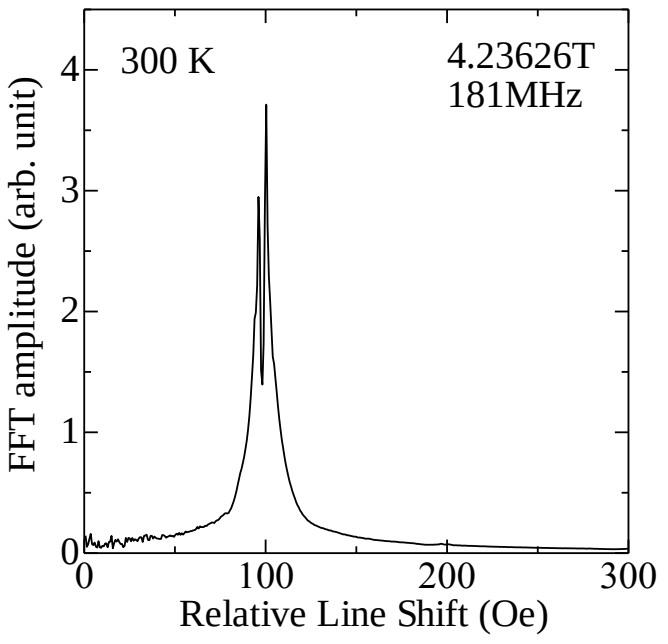

**FIG. 4** Fourier spectrum of $^1$H-NMR at 300 K.

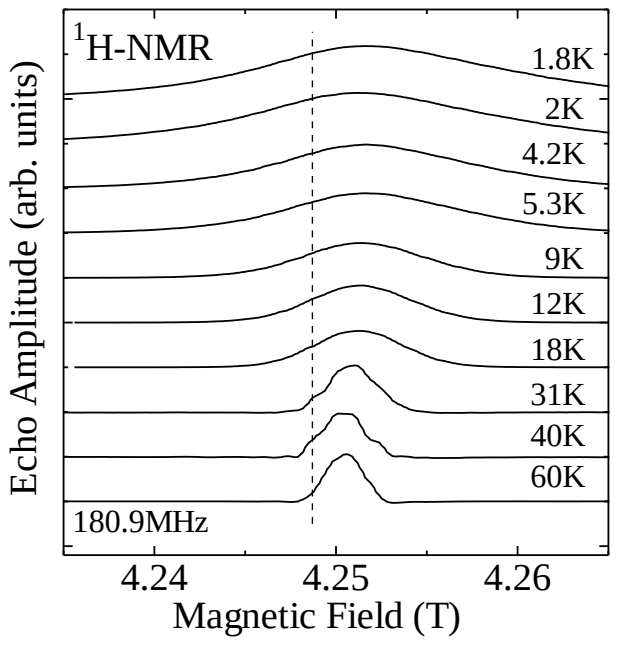

**FIG. 5** Field-swept spectra of $^1$H-NMR on Ru0/B0/GNP at various temperatures. The dashed vertical line indicates the zero-shift position.

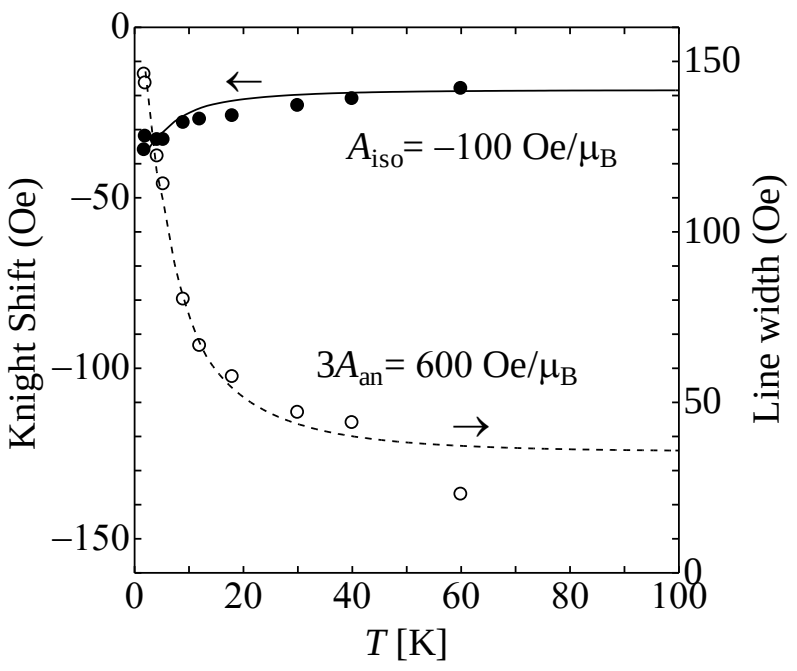

**FIG. 6** Temperature dependence of $\Delta H$, the Knight shift and $\delta H$, the line width (FWHM), scaled with that of the uniform susceptibility $\chi_0$, shown by curves. The isotropic and anisotropic parts of hyperfine coupling constants are obtained from each scaling factor.

In the entire temperature range, the spectral shape was almost symmetric, rather than the asymmetric powder pattern [14], reflecting the fact that there exist many crystallographically-inequivalent proton sites.

Both the resonance shift $\Delta H$ and the peak width $\delta H$ increased with decreasing temperature, and at 1.8 K, the latter tended to be 150 Oe FWHM. Their temperature dependence is shown in Fig. 6., where $\Delta H$ and $\delta H$ scaled well with the uniform susceptibility, giving the hyperfine coupling constants of the isotropic part $A_{iso}$ and anisotropic part $3A_{an}$ to be −100 and 600 Oe/$\mu_B$, respectively [12,13]. These moderately large values assure that the $^1$H-NMR well probes the magnetization of the entire system. The negative value of $A_{iso}$ indicates that the observed constant term in $\Delta H$, approximately −20 Oe, in the Knight shift is consistent with the positive $\chi_0$ for the Ru0/B0/GNP sample.

Finally, we discuss the origin of the paramagnetism or in other words, the positive $\chi_0$ as observed in all the three samples as shown in Table 1. As stated above, the metal gold has been known as diamagnetic matter as $\chi_0 = -27 \times 10^{-6}$ (emu/mole) [15], which is explained in terms of the effects of ionic core and band structure. The observed positive $\chi_0$ suggests an anomalous magnetic state for either GNP itself or assembled complexes. The effect of Ru0 and B0 themselves were confirmed to be negligibly small by the separate measurements of their uniform susceptibility. By fitting the same formula, we obtained $\chi_0$ for Ru0 and B0 as $+5.6 \times 10^{-7}$ and $-8.1 \times 10^{-7}$ (emu/g), respectively. Their contribution to the entire system is negligibly small, because the mass of Ru0's ($M$ =398.4) and B0's ($M$ =137.93) assembled on a GNP is below one percent of a GNP itself. Note that this estimation holds even if one assumes the assembling number of Ru0's and B0's to be a thousand, which is much larger than $n_{Ru}$, the above estimation.

The quantum size effect on nano-size metallic particles has long been studied from 80's [16], and it has been shown that the energy levels of conduction electrons become discrete at low temperatures [17], and that the Pauli paramagnetism is anomalously modified when $k_B T$ underrun $\delta E$, the average level spacing of conduction electrons in a particle [9]. This spacing $\delta E$ is roughly given as $E_F/N$, where $E_F$ and $N$ are the Fermi energy and the number of valence electrons in a particle. For the present case, $\delta E$ of GNP with the diameter 11.5 nm is estimated to be 1.4 K, which is by far much lower than the experimental temperature range, where the positive $\chi_0$ was observed in the present study. This means that its origin must be sought in other mechanisms, such as the surface gold atoms coupled with sulfur atoms in Ru0 or B0. In order to settle the problem, still more accumulation of data is necessary, such as the measurement on GNP's with different number of complexes, which is now under progress.

## IV. Conclusion

We have investigated the magnetic properties of the GNP samples, on which Ru0 complexes and B0 molecules are assembled. The average number of Ru0 complexes on each GNP (the average diameter 11.5 nm) was estimated to be 118, which is only 23 percent of the calculation based on the closed-packed structure. From the NMR spectra, it has been shown that the molecules on the surface of GNP are wavering at room temperature. The uniform susceptibility of all the samples has shown that it includes the positive constant term, which is apparently different from bulk Au, and further investigation is necessary to clarify its microscopic origin.

## Acknowledgment

This work was supported by JSPS KAKENHI Grant Number 18K03548. A part of this work was performed under the Inter-university Cooperative Research Program of the Institute for Materials Research, Tohoku University.